\documentclass[twocolumn,showpacs,preprintnumbers,amsmath,amssymb]{revtex4}

\usepackage{graphicx}
\usepackage{dcolumn}
\usepackage{bm}
\usepackage{amssymb}
\usepackage{amsfonts}
\usepackage{amsmath}
\usepackage{latexsym}
\usepackage{dsfont}
\usepackage{multirow}
\usepackage{color}
\usepackage[mathscr]{eucal}

\definecolor{nv}{rgb}{0.1,0.1,0.6}
\definecolor{pr}{rgb}{0.2,0.1,0.5}
\definecolor{mg}{rgb}{0.4,0.0,0.4}

\newcommand{\nn}{\nonumber}

\newcommand{\beq}{\begin{equation}}
\newcommand{\eeq}{\end{equation}}
\newcommand{\beqy}{\begin{eqnarray}}
\newcommand{\eeqy}{\end{eqnarray}}
\newcommand{\beqyn}{\begin{eqnarray*}}
\newcommand{\eeqyn}{\end{eqnarray*}}

\newcommand{\bs}{\begin{slide}}
\newcommand{\es}{\end{slide}}
\newcommand{\bc}{\begin{center}}
\newcommand{\ec}{\end{center}}
\newcommand{\bmin}{\begin{minipage}}
\newcommand{\emin}{\end{minipage}}

\newcommand{\bi}{\begin{itemize}}
\newcommand{\ei}{\end{itemize}}

\newcommand{\bea}{\begin{eqnarray}}
\newcommand{\eea}{\end{eqnarray}}
\newcommand{\be}{\begin{equation}}
\newcommand{\ee}{\end{equation}}

\newcommand{\ud}{\mathrm{d}}

\newcommand{\barpsi}{\overline{\psi}}

\newlength\savedwidth

\newcommand\whline{\noalign{\global\savedwidth\arrayrulewidth
\global\arrayrulewidth 1pt}%
\hline
\noalign{\global\arrayrulewidth\savedwidth}}
\newcommand{\uvec}[1]{\boldsymbol{#1}}

\newcommand{\LRD}{\overset{\leftrightarrow}{D}\!\!\!\!\!\phantom{D}}

\begin{document}


\title{Quark transverse spin-orbit correlations}

\author{Amit Bhoonah}
\email{bhoonahp@student.ethz.ch}
\affiliation{ETH Z\"urich,
R\"amistrasse 101, 8092 Z\"urich, Switzerland}

\author{C\'edric Lorc\'e}
\email{cedric.lorce@polytechnique.edu}
\affiliation{Centre de Physique Th\'eorique, \'Ecole polytechnique, 
	CNRS, Universit\'e Paris-Saclay, F-91128 Palaiseau, France}


\begin{abstract}
We extend the study of quark spin-orbit correlations in the nucleon to the case of transverse polarization. At the leading-twist level, this completes the spin structure of the quark kinetic energy-momentum tensor. In particular, we revisit the transversity decomposition of angular momentum proposed a decade ago by Burkardt and introduce a new transverse correlation, namely between quark transversity and orbital angular momentum. We also provide for the first time the Wandzura-Wilczek expression for the second Mellin moment of twist-3 transversity generalized parton distributions, along with a new sum rule. Based on lattice calculation results, we conclude that the quark transverse spin-orbit correlation is negative for both up and down flavors, just like in the longitudinal case.
\end{abstract}

\maketitle

\section{Introduction}

Understanding the nucleon spin structure is one of the key questions in hadronic physics. It opens a window on a wide range of non-pertubative effects in quantum chromodynamics (QCD) currently studied at many facilities such as Jefferson Lab, RHIC and COMPASS~\cite{Bacchetta:2016ccz}, and is a major pillar of the physics case of the future electron-ion collider (EIC)~\cite{Accardi:2012qut}. Although the proper decomposition of the nucleon spin into quark and gluon contributions constitutes one of the fundamental motivations in this field, see e.g.~\cite{Leader:2013jra,Wakamatsu:2014zza,Liu:2015xha}, the spin structure turns out to be much richer owing to spin-orbit correlations~\cite{Lorce:2011kd,Lorce:2014mxa,Lorce:2015sqe}.

In a former paper~\cite{Lorce:2014mxa}, the quark longitudinal spin-orbit correlation was studied in detail by performing a (chiral-even) helicity decomposition of the quark energy-momentum tensor. It has, in particular, been shown that the quark longitudinal spin-orbit correlation can quantitatively be expressed in terms of parton distributions. Both current phenomenological extractions based on experimental data and lattice calculations indicate that the quark spin is, in average, opposite to the quark kinetic orbital angular momentum (OAM).

In this Letter, we discuss the quark transverse spin-orbit correlation by revisiting the (chiral-odd) transversity decomposition of the quark energy-momentum tensor considered a decade ago by Burkardt~\cite{Burkardt:2005hp,Burkardt:2006ev}. Mimicking the approach used by Ji to relate angular momentum contributions to generalized parton distributions (GPDs)~\cite{Ji:1996ek}, Burkardt decomposed the symmetric energy-momentum tensor and introduced accordingly the correlation between quark transversity and total angular momentum. Here we consider the more general asymmetric energy-momentum tensor leading to another transverse correlation, now between quark transversity and OAM.

The Letter is organized as follows: In section~\ref{sec2}, we define the quark transverse spin-orbit correlation operator and express the corresponding expectation value in terms of tensor generalized form factors. In section~\ref{sec3} we relate these generalized form factors to moments of measurable parton distributions and derive for the first time the Wandzura-Wilczek expression for the second Mellin moment of twist-3 transversity generalized parton distributions, along with a new sum rule. In section~\ref{sec4}, we compare the various contributions obtained on the lattice with relativistic quark model predictions, provide an estimate of the quark transverse spin-orbit correlation, and we conclude the paper with section~\ref{sec5}.

\section{Quark spin-orbit correlations}\label{sec2}

\subsection{Decomposition based on polarization}

It is well known that the quark field operator can be decomposed into right- and left-handed contributions
\begin{equation}
\psi=\psi_R+\psi_L,\qquad \psi_{R,L}=\tfrac{1}{2}(\mathds 1\pm\gamma_5)\psi.
\end{equation}
The quark number and helicity light-front operators can then respectively be seen as the \emph{sum} and \emph{difference}
\begin{align}
\int\ud^3x\,\barpsi \gamma^+\psi&=\hat N^q_R+\hat N^q_L,\\
\int\ud^3x\,\barpsi \gamma^+\gamma_5\psi&=\hat N^q_R-\hat N^q_L
\end{align}
 of the right and left-handed densities
\begin{equation} 
\hat N^q_{R,L}=\int\ud^3x\,\barpsi_{R,L} \gamma^+\psi_{R,L},
\end{equation}
where $a^\pm=\tfrac{1}{\sqrt{2}}(a^0\pm a^3)$ for a generic four-vector $a$, and $\ud^3x=\ud x^-\,\ud^2x_\perp$.

Alternatively, the quark field operator can be decomposed into up and down transverse polarizations~\cite{Barone:2001sp}
\begin{equation}
\psi=\psi_\uparrow+\psi_\downarrow,\qquad \psi_{\uparrow,\downarrow}=\tfrac{1}{2}(\mathds 1\pm\gamma^j\gamma_5)\psi
\end{equation}
with $j=1$ or $2$. While the sum of up and down densities naturally gives the quark number operator, their difference defines the so-called quark transversity
\begin{align}
\int\ud^3x\,\barpsi \gamma^+\psi&=\hat N^q_\uparrow+\hat N^q_\downarrow,\\
\int\ud^3x\,\barpsi i\sigma^{j+}\gamma_5\psi&=\hat N^q_\uparrow-\hat N^q_\downarrow,
\end{align}
where
\begin{equation} 
\hat N^q_{\uparrow,\downarrow}=\int\ud^3x\,\barpsi_{\uparrow,\downarrow} \gamma^+\psi_{\uparrow,\downarrow}.
\end{equation}

The same decompositions can be performed with the quark light-front OAM operator
\begin{equation}
\begin{split}
\sqrt{2}\,\epsilon^\mu_{\phantom{\mu}+\alpha\beta}\int\ud^3x\,\barpsi \gamma^+x^\alpha\tfrac{i}{2}\LRD^\beta\psi&=\hat L^{q,\mu}_R+\hat L^{q,\mu}_L \\
&=\hat L^{q,\mu}_\uparrow+\hat L^{q,\mu}_\downarrow,
\end{split}
\end{equation}
where
\begin{equation}
\hat L^{q,\mu}_a=\sqrt{2}\,\epsilon^\mu_{\phantom{\mu}+\alpha\beta}\int\ud^3x\,\barpsi_a \gamma^+x^{\alpha}\tfrac{i}{2}\LRD^{\beta}\psi_a 
\end{equation}
with the convention $\epsilon_{0123}=+1$, $a=R,L,\uparrow,\downarrow$, and $\LRD^\beta=\overset{\rightarrow}{\partial}\!\!\!\!\phantom{\partial}^\beta-\overset{\leftarrow}{\partial}\!\!\!\!\phantom{\partial}^\beta-2igA^\beta$ the symmetric gauge covariant derivative. Considering instead the differences of densities leads us to longitudinal and transverse spin-orbit correlations ($\epsilon^{12}_T=-\epsilon^{21}_T=+1$ and $a^{[\mu}b^{\nu]}=a^\mu b^\nu-a^\nu b^\mu$)
\begin{align}
\hat C^q_z&\equiv\epsilon^{lk}_T\int\ud^3x\,\barpsi \gamma^+\gamma_5\,x^l \tfrac{i}{2}\LRD^k\psi=\hat L^{q,+}_R-\hat L^{q,+}_L,\label{Cz}\\
\hat C^q_j&\equiv\sqrt{2}\,\epsilon^{jl}_T\int\ud^3x\,\barpsi i\sigma^{j+}\gamma_5\,x^{[-} \tfrac{i}{2}\LRD^{l]}\psi=\hat L^{q,j}_\uparrow-\hat L^{q,j}_\downarrow\label{CT}
\end{align}
without summation over $j$ in~\eqref{CT}. These are the diagonal components of a $3\times 3$ matrix whose entries are the directions of quark polarization and OAM. 
\newline

The longitudinal spin-orbit correlation~\eqref{Cz} has been studied in~\cite{Lorce:2014mxa}. In this Letter, we focus on the transverse spin-orbit correlation~\eqref{CT} which can conveniently be rewritten as (once again without summation over $j$)
\begin{equation}\label{OAMT}
\hat C^q_j=\sqrt{2}\,\epsilon^{jl}_T\int\ud^3x \left[x^-\hat T^{j+l}_{q5}-x^l\hat T^{j+-}_{q5}\right]
\end{equation}
with $\hat T^{\lambda\mu\nu}_{q5}$ the quark energy-momentum tensor where $\gamma^\mu$ has been replaced by $i\sigma^{\lambda\mu}\gamma_5$
\begin{equation}
\hat T^{\lambda\mu\nu}_{q5}(x)=\barpsi(x) i\sigma^{\lambda\mu}\gamma_5\,\tfrac{i}{2}\LRD^\nu\psi(x).
\end{equation}
We added the index $5$ to indicate the presence of the matrix $\gamma_5$ and to distinguish it from $\hat T^{\lambda\mu\nu}_{q}=\barpsi i\sigma^{\lambda\mu}\,\tfrac{i}{2}\LRD^\nu\psi$. These two operators are equivalent owing to the identity $i\sigma^{\mu\nu}\gamma_5=\tfrac{1}{2}\,\epsilon^{\mu\nu\alpha\beta}\sigma_{\alpha\beta}$.

\subsection{Parametrization}

We find that the non-forward matrix elements of $\hat T^{\lambda\mu\nu}_{q5}$ can be parametrized in terms of seven generalized form factors (GFFs)
\beq\label{EMTparam}
\langle p',\uvec s'|\hat T^{\lambda\mu\nu}_{q5}(0)|p,\uvec s\rangle=\overline u(p',\uvec s')\Gamma^{\lambda\mu\nu}_{q5}u(p,\uvec s)
\eeq
with
\begin{align}
\Gamma^{\lambda\mu\nu}_{q5}&=\tfrac{P^\nu P^{[\lambda}\Delta^{\mu]}\gamma_5}{2M^2}\,A^q_T(t)+\tfrac{g^{\nu[\lambda}\Delta^{\mu]}\gamma_5}{2}\,\tilde A^q_T(t)\nn\\
&\quad+\tfrac{P^\nu P^{[\lambda}\gamma^{\mu]}\gamma_5}{M}\,B^q_T(t)+M\,g^{\nu[\lambda}\gamma^{\mu]}\gamma_5\,\tilde B^q_T(t)\nn\\
&\quad+\tfrac{\Delta^\nu \Delta^{[\lambda}\gamma^{\mu]}\gamma_5}{4M}\,C^q_T(t)+P^\nu i\sigma^{\lambda\mu}\gamma_5\,D^q_T(t)\nn\\
&\quad+\tfrac{P^{[\lambda} i\sigma^{\mu\nu]}\gamma_5}{2}\,\tilde D^q_T(t),\label{param}
\end{align}
where $\uvec s$ and $\uvec s'$ are the initial and final rest-frame polarization unit vectors, $M$ is the nucleon mass, $P=\tfrac{p'+p}{2}$ is the average four-momentum, and $t=\Delta^2$ is the square of the four-momentum transfer $\Delta=p'-p$. Note that the last term is totally antisymmetric over all three Lorentz indices, so that $\tfrac{P^{[\lambda} i\sigma^{\mu\nu]}\gamma_5}{2}=P^\nu i\sigma^{\lambda\mu}\gamma_5+P^{[\lambda} i\sigma^{\mu]\nu}\gamma_5$. To recover the twist-2 parametrization of  H\"agler and Diehl~\cite{Hagler:2004yt,Diehl:2005jf}, one has to symmetrize over the pair of indices $\{\mu\nu\}$, antisymmetrize over the pair of indices $[\lambda\mu]$ and remove all the traces~\cite{Geyer:1999uq}. As a result, the tilde GFFs become redundant
\begin{align}
3\tilde A^q_T(t)&\stackrel{\text{tw}-2}{=}\left(\tau-1\right)A^q_T(t)+C^q_T(t)-D^q_T(t),\label{LTAtT}\\
3\tilde B^q_T(t)&\stackrel{\text{tw}-2}{=}\left(\tau-1\right)B^q_T(t)-\tau C^q_T(t)+D^q_T(t),\\
3\tilde D^q_T(t)&\stackrel{\text{tw}-2}{=}-D^q_T(t),\label{LTDtT}
\end{align}
where $\tau=\tfrac{t}{4M^2}$. This means that only four GFFs survive at leading twist in agreement with the results of~\cite{Hagler:2004yt,Diehl:2005jf}. More precisely, we find that the two parametrizations at leading twist are related as follows
\begin{align}
A^q_T(t)+B^q_T(t)&=B_{T20}(t),\\
B^q_T(t)&=2\tilde A_{T20}(t)+B_{T20}(t),\label{BTq}\\
C^q_T(t)&=2\tilde B_{T21}(t),\\
D^q_T(t)-B^q_T(t)&=A_{T20}(t)-2\tau\tilde A_{T20}(t).\label{DTq}
\end{align}

We are ultimately interested in the matrix elements of Eq.~\eqref{OAMT} which involves one explicit power of $x$. It is therefore sufficient to expand Eq.~\eqref{EMTparam} up to linear order in $\Delta$~\cite{Leader:2013jra,Bakker:2004ib}. Using the light-front spinors (see e.g. Appendix A of~\cite{Lorce:2011zta}) with the same rest-frame polarization $\uvec s'=\uvec s=(\uvec s_\perp,s_z)$, we obtain 
\begin{align}\label{expansion}
\langle p'&,\uvec s|\hat T^{\lambda\mu\nu}_{q5}|p,\uvec s\rangle=\nn\\
&\left[\tfrac{2P^\nu P^{[\lambda}S^{\mu]}}{M}+\tfrac{MP^\nu\,i\epsilon^{+\lambda\mu\Delta}}{P^+}\right]\left(B^q_T-D^q_T\right)\nn\\
&+\left[2M\,g^{\nu[\lambda}S^{\mu]}+\tfrac{M\,g^{\nu[\lambda}i\epsilon^{\mu]+P\Delta}}{P^+}\right]\tilde B^q_T\nn\\
&-\tfrac{P^\nu i\epsilon^{\lambda\mu P\Delta}}{M}\,B^q_T-M\,i\epsilon^{\lambda\mu\nu\Delta}\,\tilde D^q_T+\mathcal O(\Delta^2)
\end{align}
with the covariant spin vector $S^\mu=[s_zP^+,-s_zP^-+\tfrac{\uvec P_\perp}{P^+}\cdot(M\uvec s_\perp+\uvec P_\perp s_z),M\uvec s_\perp+\uvec P_\perp s_z]$ satisfying $P\cdot S=0$ and $S^2=-M^2(1-\tau\,s^2_z)$. For convenience, we removed the argument of the GFFs when evaluated at $t=0$, e.g. $B^q_T\equiv B^q_T(0)$, and we wrote four-vectors as indices whenever they appear contracted, e.g. $\epsilon^{\lambda\mu P\Delta}\equiv\epsilon^{\lambda\mu\alpha\beta}P_\alpha\Delta_\beta$. 

Substituting the expansion~\eqref{expansion} into the matrix element of~\eqref{OAMT} and working in the symmetric light-front frame, i.e. with $\uvec P_\perp=\uvec 0_\perp$, we find
\beq
C^q_j\equiv\tfrac{\langle P,\uvec s|\hat C^q_j|P,\uvec s\rangle}{\langle P,\uvec s|P,\uvec s\rangle}=-\tfrac{M}{2\sqrt{2}P^+}(B^q_T+2\tilde B^q_T+4\tilde D^q_T).
\eeq
Like the longitudinal one~\cite{Lorce:2014mxa}, the transverse spin-orbit correlation does not depend on the nucleon polarization, as a consequence of parity conservation. However, unlike the longitudinal case, it depends on the light-front momentum $P^+$. This dependence can be understood as coming from the transverse component of the OAM operator, since the transversity operator is leading twist and hence $P^+$ independent. Indeed, contrary to the longitudinal component, the decomposition of the transverse component of total angular momentum into spin and OAM contributions is known to be in general frame dependent~\cite{Leader:2013jra}. Note also that in the rest frame $\sqrt{2}P^+=M$ we recover the familiar $\tfrac{1}{2}$ global factor~\cite{Lorce:2014mxa,Ji:1996ek}.

\section{Link with parton distributions}\label{sec3}

No fundamental probe coupling to $\hat T^{\lambda\mu\nu}_{q5}$ is known in particle physics. It is however possible to relate the corresponding GFFs to specific moments of measurable parton distributions. From the leading-twist component $\hat T^{j++}_{q5}$, we find in agreement with~\cite{Hagler:2004yt,Diehl:2005jf}
\begin{align}
\int\ud x\, x H^q_T(x,\xi,t)&=-\tau A^q_T(t)-B^q_T(t)+D^q_T(t)\nn\\
&=A_{T20}(t),\\
\int\ud x\, x E^q_T(x,\xi,t)&=A^q_T(t)+B^q_T(t)=B_{T20}(t),\\
\int\ud x\, x \tilde H^q_T(x,\xi,t)&=-\tfrac{1}{2} A^q_T(t)=\tilde A_{T20}(t),\\
\int\ud x\, x \tilde E^q_T(x,\xi,t)&=-\xi C^q_T(t)=-2\xi\tilde B_{T21}(t),\label{xEtT}
\end{align}
where the skewness variable is given by $\xi=-\Delta^+/2P^+$ and the functions $H^q_T(x,\xi,t)$, $E^q_T(x,\xi,t)$, $\tilde H^q_T(x,\xi,t)$, and $\tilde E^q_T(x,\xi,t)$ are the GPDs parametrizing the non-local twist-2 tensor light-front quark correlator~\cite{Diehl:2001pm,Diehl:2003ny,Meissner:2009ww}
\begin{align}
\frac{1}{2}\int\frac{\ud z^-}{2\pi}\,e^{ixP^+z^-}&\langle p',\uvec s'|\barpsi(-\tfrac{z^-}{2})i\sigma^{j+}\gamma_5\mathcal W\psi(\tfrac{z^-}{2})|p,\uvec s\rangle\nn\\
&=\tfrac{i\epsilon_T^{jl}}{2P^+}\,\overline u(p',\uvec s')\Gamma^{+l}_{qT}u(p,\uvec s)
\end{align}
with $\mathcal W=\mathcal P \exp[ ig\int^{-z^-/2}_{z^-/2}\ud y^-A^+(y^-)]$ a straight light-front Wilson line and
\begin{align}
&\Gamma^{+l}_{qT}=i\sigma^{+l}\,H^q_T(x,\xi,t)+\tfrac{\gamma^+\Delta^l_\perp-\Delta^+\gamma^l_\perp}{2M}\, E^q_T(x,\xi,t)\nn\\
&+\tfrac{P^+\Delta^l_\perp}{M^2}\,\tilde H^q_T(x,\xi,t) -\tfrac{P^+\gamma^l_\perp}{M}\,\tilde E^q_T(x,\xi,t)
\end{align}
written in the symmetric frame $\uvec P_\perp=\uvec 0_\perp$.

\subsection{Equations of motion}

The relations for the tilde GFFs can be obtained using the following QCD identities
\begin{align}
\barpsi i\sigma^{\lambda\mu}\gamma_5\, i\LRD_{\mu}\psi&=2m\,\barpsi\gamma^\lambda\gamma_5\psi+i\partial^\lambda(\barpsi\gamma_5\psi),\\
\barpsi i\sigma^{[\lambda\mu}\gamma_5\, i\LRD^{\nu]}\psi&=-2\epsilon^{\lambda\mu\nu\alpha}\partial_\alpha(\barpsi\psi),\label{QCDidentity}
\end{align}
where $m$ is the quark mass. Taking the corresponding matrix elements and using some Gordon and $\epsilon$-identities, we find
\begin{align}
\left(\tau-1\right)A^q_T(t)-3\tilde A^q_T(t)&+C^q_T(t)-D^q_T(t)\nn\\
&=\tfrac{m}{M}\,G^q_P(t)-\Pi_q(t),\label{AtT}\\
\left(\tau-1\right)B^q_T(t)-3\tilde B^q_T(t)&-\tau C^q_T(t)+D^q_T(t)\nn\\
&=\tfrac{m}{M}\,G^q_A(t),\\
D^q_T(t)+3\tilde D^q_T(t)&=\Sigma_q(t),\label{DtT}
\end{align}
which generalize the leading-twist expressions~\eqref{LTAtT}-\eqref{LTDtT}. The FFs on the right-hand side parametrize the scalar, pseudoscalar and axial-vector local correlators as follows
\begin{align}
\langle p',\uvec s'|\barpsi\psi|p,\uvec s\rangle&=\overline u(p',\uvec s')\Gamma_{qS}u(p,\uvec s),\\
\langle p',\uvec s'|\barpsi\gamma_5\psi|p,\uvec s\rangle&=\overline u(p',\uvec s')\Gamma_{qP}u(p,\uvec s),\\
\langle p',\uvec s'|\barpsi \gamma^\mu\gamma_5\psi|p,\uvec s\rangle&=\overline u(p',\uvec s')\Gamma^\mu_{qA}u(p,\uvec s)
\end{align}
with
\begin{align}
\Gamma_{qS}&=\mathds 1\,\Sigma_q(t),\\
\Gamma_{qP}&=\gamma_5\,\Pi_q(t),\\
\Gamma^\mu_{qA}&=\gamma^\mu\gamma_5\, G^q_A(t)+\tfrac{\Delta^\mu\gamma_5}{2M}\,G^q_P(t).
\end{align}

The quark transverse spin-orbit correlation is therefore given by the expression
\begin{align}
\tfrac{\sqrt{2}P^+}{M}\,C^q_j&=\tfrac{1}{3}\int\ud x\,x[H^q_T(x,0,0)+\tfrac{1}{2}\bar E^q_T(x,0,0)]\nn\\
&\qquad\qquad-\tfrac{2}{3}[\Sigma_q(0)-\tfrac{m}{2M}\,G^q_A(0)],\label{Cqj}
\end{align}
where $\bar E^q_T(x,\xi,t)\equiv 2\tilde H^q_T(x,\xi,t)+E^q_T(x,\xi,t)$. Interestingly, it is very similar to the corresponding expression for the longitudinal spin-orbit correlation~\cite{Lorce:2014mxa}
\beq
C^q_z=\tfrac{1}{2}\int\ud x\,x\tilde H_q(x,0,0)-\tfrac{1}{2}\,[F^q_1(0)-\tfrac{m}{2M}\,H^q_1(0)]
\eeq
and Ji's relation~\cite{Ji:1996ek} for the quark OAM
\beq\label{Jirel}
L^q_z=\tfrac{1}{2}\int\ud x\,x[H_q(x,0,0)+E_q(x,0,0)]-\tfrac{1}{2}\,G^q_A(0).
\eeq
One might be surprised that Eq.~\eqref{Cqj} involves thirds instead of halves. They appear because of the factors 3 in Eqs.~\eqref{AtT}-\eqref{DtT} which trace back to the fact that $C^q_j$ is defined from a rank-3 tensor, while $L^q_z$ and $C^q_z$ are defined from rank-2 tensors.

Let us stress that the quark transverse spin-orbit correlation introduced in this Letter corresponds actually to the correlation between quark transversity and OAM $\langle L^q_zT^q_z\rangle$. The similarity of our result~\eqref{Cqj} with Eq.~\eqref{Jirel}, which can be understood as the difference between total angular momentum and spin $\langle L^q_zS^N_z\rangle=\langle J^q_zS^N_z\rangle-\langle S^q_zS^N_z\rangle$ according to~\cite{Ji:1996ek,Lorce:2014mxa}, hints towards the identifications $\langle J^q_xT^q_x\rangle\propto\int\ud x\,x[H^q_T(x,0,0)+\tfrac{1}{2}\bar E^q_T(x,0,0)]$ and $\langle S^q_xT^q_x\rangle\propto \Sigma_q(0)$ in the chiral limit $m=0$. In particular, it suggests that the scalar charge can be interpreted as a measure of the correlation between quark transversity and spin. This interpretation is further supported by the following simple reasoning in instant form. The difference between spin $\psi^\dagger\sigma^{ij}\psi$ and transversity $\psi^\dagger\gamma^0\sigma^{ij}\psi$ is a factor $\gamma^0$, and hence is of relativistic nature~\cite{Jaffe:1991kp}. The correlation between spin and transversity then reads $\psi^\dagger\sigma^{ij}\gamma^0\sigma^{ij}\psi$ (without summation over $i,j$), which simplifies to the scalar bilinear $\psi^\dagger\gamma^0\psi=\overline\psi\psi$.

\subsection{Twist-3 tensor GPDs}

The tilde GFFs can alternatively be expressed in terms of twist-3 tensor GPDs. From the twist-3 components $\hat T^{jl+}_{q5}$ and $\hat T^{+-+}_{q5}$, we obtain the following relations
\begin{align}
\int\ud x\,x H'^q_2&=-\xi\!\left[\tau C^q_T(t)+D^q_T(t)+\tilde D^q_T(t)\right],\label{xHp}\\
\int\ud x\,x E'^q_2&=\xi\!\left[C^q_T(t)+D^q_T(t)+\tilde D^q_T(t)\right],\label{xEp}\\
\int\ud x\,x\tilde H'^q_2&=-\left(1-\tau\right)B^q_T(t)-\tilde B^q_T(t)+D^q_T(t),\\
\int\ud x\,x\tilde E'^q_2&=\xi\!\left[\left(1-\tau\right)A^q_T(t)+\tilde A^q_T(t)+D^q_T(t)\right],
\end{align}
where the functions $H'^q_2(x,\xi,t)$, $E'^q_2(x,\xi,t)$, $\tilde H'^q_2(x,\xi,t)$, and $\tilde E'^q_2(x,\xi,t)$ are the GPDs parametrizing the non-local twist-3 tensor light-front quark correlators~\cite{Meissner:2009ww}
\begin{align}
\frac{1}{2}\int\frac{\ud z^-}{2\pi}\,e^{ixP^+z^-}&\langle p',\uvec s'|\barpsi(-\tfrac{z^-}{2})i\sigma^{jl}\gamma_5\mathcal W\psi(\tfrac{z^-}{2})|p,\uvec s\rangle\nn\\
&=\tfrac{M}{2(P^+)^2}\,\overline u(p',\uvec s')\Gamma^{jl}_{qT}u(p,\uvec s),\\
\frac{1}{2}\int\frac{\ud z^-}{2\pi}\,e^{ixP^+z^-}&\langle p',\uvec s'|\barpsi(-\tfrac{z^-}{2})i\sigma^{+-}\gamma_5\mathcal W\psi(\tfrac{z^-}{2})|p,\uvec s\rangle\nn\\
&=\tfrac{M}{2(P^+)^2}\,\overline u(p',\uvec s')\Gamma^{+-}_{qT}u(p,\uvec s)
\end{align}
with
\begin{align}
\Gamma^{jl}_{qT}&=-i\epsilon_T^{jl}\left[\gamma^+H'^q_2(x,\xi,t)+\tfrac{i\sigma^{+\Delta}}{2M}\, E'^q_2(x,\xi,t)\right],\\
\Gamma^{+-}_{qT}&=\gamma^+\gamma_5\, \tilde H'^q_2(x,\xi,t)+\tfrac{P^+\gamma_5}{M}\, \tilde E'^q_2(x,\xi,t).
\end{align}

Since the eight (twist-2 and twist-3) tensor GPD moments are expressed in terms of seven GFFs, there exists a sum rule among them. Adding Eqs.~\eqref{xHp} and~\eqref{xEp} and using Eq.~\eqref{xEtT}, we find
\begin{equation}
\int\ud x\,x\left[(1-\tau)\tilde E^q_T+H'^q_2+E'^q_2\right]=0.
\end{equation}
Moreover, using the relations~\eqref{AtT}-\eqref{DtT}, we obtain
\begin{align}
&\tfrac{1}{\tau-1}\tfrac{1}{\xi}\int\ud x\,x\left[H'^q_2+\tau E'^q_2\right]=\tfrac{1}{3}\Sigma_q\nn\\
&\qquad\qquad +\tfrac{2}{3}\int\ud x\,x\left[H^q_T-2\tau \tilde H^q_T+\bar E^q_T\right],\\
&\int\ud x\,x\,\tilde H'^q_2=\tfrac{m}{3M}G^q_A\nn\\
&\qquad\qquad +\tfrac{1}{3}\int\ud x\,x\left[2\left(H^q_T+\tau E^q_T\right)-\tfrac{\tau}{\xi}\tilde E^q_T\right],\\
&\tfrac{1}{\xi}\int\ud x\,x\,\tilde E'^q_2=-\tfrac{m}{3M}G^q_P+\tfrac{1}{3}\Pi_q\nn\\
&\qquad\qquad +\tfrac{1}{3}\int\ud x\,x\left[2\left(H^q_T+E^q_T\right)-\tfrac{1}{\xi}\tilde E^q_T\right].
\end{align}
In the massless quark limit, these expressions provide the Wandzura-Wilczek approximation to the second Mellin moment of twist-3 tensor GPDs. Note that they are exact since no three-parton correlators were involved in the derivation, similar to what was observed in the chiral-even sector~\cite{Penttinen:2000dg,Kivel:2000cn}.

Thanks to these results, the quark transverse spin-orbit correlation can now be written in terms of tensor GPDs only
\begin{align}
\tfrac{\sqrt{2}P^+}{M}\,C^q_j&=\int\ud x\,x[H^q_T(x,0,0)+\tfrac{3}{2}\bar E^q_T(x,0,0)]\nn\\
&+\int\ud x\,x[\tilde H'^q_2(x,0,0)+2H'^q_{2\xi}(x,0,0)],
\end{align}
where $H'^q_{2\xi}(x,0,t)\equiv\lim_{\xi\to 0}\tfrac{1}{\xi}H'^q_2(x,\xi,t)$. It is the chiral-odd analogue of the Penttinen-Polyakov-Shuvaev-Strikman relation for the Ji or kinetic OAM~\cite{Kiptily:2002nx,Penttinen:2000dg}
\beq
L^q_z=-\int\ud x\,xG^q_2(x,0,0)
\eeq
and of the relation we found in~\cite{Lorce:2014mxa} for the quark longitudinal spin-orbit correlation
\beq
C^q_z=-\int\ud x\,x[\tilde G^q_2(x,0,0)+2\tilde G^q_4(x,0,0)].
\eeq
Note that this time both twist-2 and twist-3 GPDs are necessary to express the quark transverse spin-orbit correlation. This may be due to the fact that transversity does not coincide with transverse spin~\cite{Barone:2001sp}.

\section{Discussion}\label{sec4}

\subsection{Burkardt's correlation}

In the former sections, we worked with the asymmetric quark kinetic energy-momentum tensor and performed a decomposition in terms of quark transversity states. The quark transverse spin-orbit correlation $C^q_x=\langle L^q_xT^q_x\rangle$ can therefore alternatively be seen as the transversity asymmetry of the quark OAM, i.e. $\langle\delta^x L^x_q\rangle$ following Burkardt's notation.

This has to be contrasted with the work of Burkardt in~\cite{Burkardt:2005hp,Burkardt:2006ev} which is based on the Belinfante or symmetric quark kinetic energy-momentum tensor~\cite{Belinfante:1939,Belinfante:1940,Rosenfeld:1940}. Since in this case the total angular momentum assumes a purely orbital form, Burkardt interpreted his correlation as the transversity asymmetry of the quark total angular momentum $\langle\delta^x J^x_q\rangle$. It may be tempting to identify it with the correlation between quark transversity and total angular momentum $\langle J^q_xT^q_x\rangle$, just like we identified the quark transverse spin-orbit correlation $C^q _x=\langle L^q_xT^q_x\rangle$ with the transversity asymmetry of the quark OAM $\langle\delta^xL^x_q\rangle$. This is, however, not consistent since $\langle T^{\lambda+\nu}_{q5}\rangle\neq\langle\tfrac{1}{2}\,T^{\lambda\{+\nu\}}_{q5}\rangle$ as one can see from the QCD identity~\eqref{QCDidentity}, whereas we have $\langle T^{+\nu}_{q}\rangle=\langle\tfrac{1}{2}\,T^{\{+\nu\}}_{q}\rangle$ for the unpolarized quark energy-momentum tensor. In other words, symmetrization and transversity decomposition are not compatible, so that we expect in general $\langle\delta^xJ^x_q\rangle\neq \langle\delta^xL^x_q\rangle+\langle\delta^xS^x_q\rangle$.

In the light-front formalism, Burkardt's quark operator is given by 
\begin{equation}\label{deltaJop}
\hat{\mathds C}^q_x=\sqrt{2}\int\ud^3x\left[x^-\tfrac{1}{2}\hat T^{1\{+2\}}_{q5}-x^2\tfrac{1}{2}\hat T^{1\{+-\}}_{q5}\right]
\end{equation}
which is like our operator $\hat C^q_x$ time-dependent \footnote{We would like to correct a remark in~\cite{Lorce:2014mxa}. The quark longitudinal spin-orbit operator $\hat{\mathds C}^q_z$ is actually not time-independent since the operator $\hat T^{\{\mu\nu\}}_{q5}$ is in general not conserved. This has however no practical consequences since we are only interested in matrix elements where the initial and final energies are the same.}.  Symmetrizing the expansion~\eqref{expansion} over the pair of indices $\{\mu\nu\}$, we find
\begin{equation}\label{deltaJLF}
\langle\delta^x\!J^x_q\rangle\equiv\tfrac{\langle P,\uvec s|\hat{\mathds C}^q_x|P,\uvec s\rangle}{\langle P,\uvec s|P,\uvec s\rangle}=-\tfrac{M}{2\sqrt{2}P^+}(B^q_T+2\tilde B^q_T-2 D^q_T).
\end{equation}
Note that the GFF $\tilde D^q_T$ naturally drops out of the final result since it is associated with a Lorentz structure antisymmetric in the pair of indices $[\mu\nu]$. 

Actually, Burkardt used the instant-form (IF) formalism, where the quark operator is defined as
\begin{equation}
\hat{\mathds C}^q_{x,IF}=\int\ud^3x\,(x^2\tfrac{1}{2}\hat T^{1\{03\}}_{q5}-x^3\tfrac{1}{2}\hat T^{1\{02\}}_{q5}).
\end{equation}
We obtain in this case (once again with $\uvec P_\perp=\uvec 0_\perp$)
\begin{equation}\label{deltaJIF}
\langle\delta^x\!J^x_q\rangle_{IF}=\tfrac{1}{2}\!\left(\tfrac{E-M}{M}\,B^q_T+D^q_T\right)
\end{equation}
which is reminiscent of Leader's result for the transverse Belinfante angular momentum~\cite{Leader:2011cr,Leader:2013jra}
\begin{equation}
\langle J^x_q\rangle_{IF}=\tfrac{1}{2}\left[\tfrac{E-M}{M}\,B^q+(A^q+B^q)\right].
\end{equation}
In the rest frame, we recover Burkardt's result
\begin{equation}\label{Burkres}
\langle\delta^x\!J^x_q\rangle_{IF,\text{rest}}=\tfrac{1}{2}D^q_T=\tfrac{1}{2}(A_{T20}+2\tilde A_{T20}+B_{T20}),
\end{equation}
where we have used Eqs.~\eqref{BTq} and~\eqref{DTq}.

At first sight, it may seem odd that the light-front and instant-form results~\eqref{deltaJLF} and~\eqref{deltaJIF} have different high-energy limits. This is because the transverse OAM light-front operator involves the $a^-$ component, whereas the instant-form operator involves $a^3=\tfrac{1}{\sqrt{2}}(a^+-a^-)$. Therefore, in the high-energy limit $E\gg M$, the light-front result behaves as $\mathcal O(E^{-1})$ whereas the instant-form result behaves as $O(E)$. In other words, the instant-form operator contains contributions which are of higher-twist compared to the corresponding light-front operator.

Beside the instant-form approach, Burkardt proposed in~\cite{Burkardt:2005hp} a heuristic derivation of Eq.~\eqref{Burkres} based on the light-front operator $\hat T^{j++}_{q5}$, allowing for an intuitive partonic interpretation in impact-parameter space. Considering the matrix element of the operator $\sqrt{2}\int\ud^3x\,x^2\hat T^{1++}_{q5}$, one obtains $\tfrac{P^+}{\sqrt{2}M}\,B^q_T$ which coincides, as expected, with the instant form result~\eqref{deltaJIF} in the infinite-momentum frame. Working in the rest frame to invoke rotational symmetry, Burkardt added an extra term $\tfrac{1}{2}\int\ud x\,xH^q_T(x,0,0)=\tfrac{1}{2}(D^q_T-B^q_T)$ to account for an overall transverse displacement of the center of light-front momentum with respect to the origin, a relativistic effect associated with rotating bodies. Although quite appealing, this interpretation is however not satisfactory since, as stressed in~\cite{Leader:2012md,Harindranath:2012wn,Liu:2015xha}, the term $\int\ud^3x\,x^j\hat T^{++}_q$ is part of the transverse light-front boost operator and not the transverse light-front rotation operator.

\subsection{Estimates from lattice calculations}

In order to determine the quark transverse spin-orbit correlation $C^q_x$, we need to know four quantities given in Eq.~\eqref{Cqj}. In practice, we can neglect the contribution of the axial FF since it appears multiplied by the mass ratio $m/3M\approx 10^{-3}$ for $u$ and $d$ quarks. 

\begin{table*}[t!]
\begin{center}
\caption{\footnotesize{Predictions for the scalar charges $\Sigma_q$, tensor charges $\delta q=\int\ud x\,H^q_T(x,0,0)$, anomalous tensor charges $\kappa^q_T=\int\ud x\,\bar E^q_T(x,0,0)$, and second Mellin moments of $H^q_T(x,0,0)$ and $\bar E^q_T(x,0,0)$ for $q=u,d$ from the light-front constituent quark model (LFCQM) and the light-front chiral quark-soliton model (LF$\chi$QSM) at the scale $\mu^2\sim 0.26$ GeV$^2$, and from lattice calculations at the scale $\mu^2=4$ GeV$^2$.}}\label{Modelresults}
\begin{tabular}{c@{\quad}|@{\quad}c@{\quad}c@{\quad}|@{\quad}c@{\quad}c}\whline
&\multicolumn{2}{c@{\quad}|@{\quad}}{Quark model}&\multicolumn{2}{c@{\quad}}{Lattice}\\
&LFCQM&LF$\chi$QSM&QCDSF/UKQCD Coll.&Abdel-Rehim \emph{et al.}\vspace{-.3cm}\\
&\cite{Lorce:2011kd,Pasquini:2007xz,Lorce:2011dv}&\cite{Lorce:2011kd,Lorce:2007fa,Lorce:2011dv}&\cite{Gockeler:2005cj,Gockeler:2006zu}&\cite{Abdel-Rehim:2015owa}\\
\whline

$\delta u$&$1.165$&$1.241$&$0.857(13)$&$0.791(53)$\\
$\delta d$&$-0.291$&$-0.310$&$-0.212(5)$&$-0.236(33)$\\
$\kappa^u_T$&$3.98$&$3.83$&$2.93(13)$&--\\
$\kappa^d_T$&$2.60$&$2.58$&$1.90(9)$&--\\
\hline
$\int\ud x\,xH^u_T$&$0.395$&$0.418$&$0.268(6)$&$0.264(25)$\\
$\int\ud x\,xH^d_T$&$-0.099$&$-0.105$&$-0.052(2)$&$-0.045(21)$\\
$\int\ud x\,x\bar E^u_T$&$1.080$&$1.072$&$0.420(31)$&--\\
$\int\ud x\,x\bar E^d_T$&$0.737$&$0.748$&$0.260(23)$&--\\
\hline
$\Sigma_{u+d}$&--&--&--&$8.93(86)$\\
$\Sigma_{u-d}$&--&--&--&$2.20(54)$\\
\whline
\end{tabular}
\end{center}
\end{table*}

So far, the second Mellin moment of quark transversity GPDs have not yet been extracted from experimental data. We will therefore rely on lattice QCD calculations. In table~\ref{Modelresults} we summarize the results obtained by the QCDSF/UKQCD Collaboration~\cite{Gockeler:2005cj,Gockeler:2006zu} for the lowest two Mellin moments of the tensor GPDs $H^q_T$ and $E^q_T$. They are in very good agreement with a more recent calculation by Abdel-Rehim \emph{et al.}~\cite{Abdel-Rehim:2015owa} which also provides us with an estimate of the quark scalar charges. These values are compared with the predictions of two relativistic quark models, namely the light-front constituent quark model (LFCQM) and the light-front chiral quark-soliton model (LF$\chi$QSM)~\cite{Lorce:2011kd,Pasquini:2007xz,Lorce:2011dv,Lorce:2007fa}. Note that the second Mellin moments are new results we obtained within these models. Even though the lattice and quark model results correspond to two different scales, namely $\mu^2=4$ GeV$^2$ and $\mu^2\sim 0.26$ GeV$^2$, we observe that they are in qualitative agreement.

Using the lattice results from table~\ref{Modelresults} in Eq.~\eqref{Cqj}, we get 
\begin{equation}
C^u_x\approx -3.6,\qquad C^d_x\approx -2.2.
\end{equation}
These numbers can be compared with the values for the longitudinal spin-orbit correlation $C^u_z\approx -0.9$ and $C^d_z\approx -0.53$ we obtained in~\cite{Lorce:2014mxa}. In both cases we found negative spin-orbit correlations, meaning that the quark polarization and kinetic OAM are, in average, anti-correlated. In the rest frame, one can expect from spherical symmetry that longitudinal and transverse spin-orbit correlations should be equal. This does not contradict our results because we considered the correlation of quark OAM with transversity and not with transverse spin. Note that the large numbers we obtained for the transverse spin-orbit correlation are mainly driven by the scalar charges, just like the longitudinal spin-orbit correlation is mainly driven by the vector charges.

The numbers in table~\ref{Modelresults} can also be used to estimate Burkardt's correlation. In this case, we obtain from Eq.~\eqref{Burkres}
\begin{align}
\langle\delta^xJ^x_u\rangle_\text{latt.}&=0.344,&\langle\delta^xJ^x_d\rangle_\text{latt.}&=0.104,\\
\langle\delta^xJ^x_u\rangle_\text{LFCQM}&=0.737,& \langle\delta^xJ^x_d\rangle_\text{LFCQM}&=0.319,\\
\langle\delta^xJ^x_u\rangle_\text{LF$\chi$QSM}&=0.745,& \langle\delta^xJ^x_d\rangle_\text{LF$\chi$QSM}&=0.321,
\intertext{which can be compared to the values obtained in~\cite{Pasquini:2005dk}}
\langle\delta^xJ^x_u\rangle_\text{HYP}&=0.39,& \langle\delta^xJ^x_d\rangle_\text{HYP}&=0.10,\\
\langle\delta^xJ^x_u\rangle_\text{HO}&=0.68,& \langle\delta^xJ^x_d\rangle_\text{HO}&=0.28,
\end{align}
for the hypercentral (HYP) and harmonic oscillator (HO) models.

\section{Conclusions}\label{sec5}

We introduced and discussed the quark transverse-spin orbit correlation, which is a new piece of information characterizing the nucleon spin structure. We showed that this correlation can be expressed in terms of tensor generalized parton distributions, scalar charges and axial-vector charges. Using results from lattice QCD calculations, we concluded that the quark transverse spin-orbit correlation is very likely negative, just like its longitudinal counterpart. In other words, it is expected that the quark kinetic orbital angular momentum is in average opposite to the quark spin.

In the process, we compared our quark transverse-spin orbit correlation with Burkardt's transverse correlation, and obtained several other interesting results. We derived a new sum rule relating twist-2 and twist-3 transversity generalized parton distributions, and also obtained the Wandzura expression for the second Mellin moment of twist-3 transversity generalized parton distributions, which is exact in the chiral limit like in the chiral-even sector. Finally, comparing Ji's expression for quark kinetic orbital angular momentum to our expression for the quark transverse spin-orbit correlation, we suggested that the scalar charge could be interpreted as a measure of the correlation between quark transversity and transverse spin.

\end{document}